\documentclass{jps-cp}
\usepackage{txfonts} 
 \usepackage{amsmath}
 \usepackage{multicol}
\usepackage{wrapfig}

\title{
High-Field Ultrasonic Study of CeIrIn$_5$
}

\author{
Ryosuke \textsc{Kurihara}$^{1, 2}$, 
Atsushi \textsc{Miyake}$^{2}$,
Ryoma \textsc{Tsunoda}$^{3}$,
Yusuke \textsc{Hirose}$^{3}$,
Rikio \textsc{Settai}$^{3}$,
and
Masashi \textsc{Tokunaga}$^{2}$
}

\inst{
$^1$Department of Physics, Faculty of Science and Technology, Tokyo University of Science, Noda, Chiba 278-8510, Japan	\\
$^2$The Institute for Solid State Physics, The University of Tokyo, Kashiwa, Chiba 277-8581, Japan	\\
$^3$Department of Physics, Niigata University, Niigata 950-2181, Japan	\\
}

\email{
r.kurihara@rs.tus.ac.jp or iron.pnictide.man@gmail.com
}

\recdate{
July 15, 2022
}

\abst{
We performed ultrasonic and temperature measurements under pulsed magnetic fields on CeIrIn$_5$ to investigate whether or not the origin of metamagnetic transition and field-induced Lifshitz transition was described by the anisotropic electronic ordering with crystal symmetry breaking.
We observed elastic anomalies with the negative magnetocaloric effects around the field-induced Lifshitz transition and the metamagnetic transition.
On the other hand, the elastic anomalies appeared both in the transverse elastic constants $(C_{11} - C_{12})/2$ and $C_{66}$ in CeIrIn$_5$ while $(C_{11} - C_{12})/2$ showed significant elastic anomaly at the crystal symmetry breaking fields in CeRhIn$_5$.
These results indicate that the anisotropic electronic ordering with crystal symmetry breaking is absent in high-magnetic fields in CeIrIn$_{5}$.
}

\kword{
CeIrIn$_5$,
Lifshitz transition,
metamagnetic transition,
electronic nematic state,
ultrasonic measurement,
high-magnetic field
}

\begin{document}
\maketitle

\section{Introduction}

Spontaneous symmetry breaking characterized by anisotropic electronic states has intensively been investigated in recent solid-state physics.
One of the conventional examples of such orderings can be the electric multipole ordering in the localized $4f$-electron systems.
Symmetry breaking due to the ferro-type ordering of an electric quadrupole causes the anisotropic electronic states described by the difference in the occupation number of electronic states.
As a result, a crystal system also exhibits symmetry breaking characterized by the lowering of space group via electron-phonon interaction.
In this case, the response of multipole as a function of temperature is described by the multipole susceptibility based on the wave functions of $4f$ electrons under the crystalline electric field (CEF).
Such an ordering has also been discussed in $3d$-, $4d$-, and $5f$-electron systems.

On the other hand, the field-induced (FI) anisotropic electronic ordering
\cite{Ronning_Nature313},
which is called electronic nematic (EN) ordering, has been observed in CeRhIn$_5$ with localized $4f$ electrons in zero magnetic fields.
Since $[100]$ ($[110]$) crystallographic orientation is equal to $[010]$ ($[1\bar{1}0]$) in CeRhIn$_5$ with the space group $P4/mmm$ ($D_{4h}^1$)
\cite{Hegger_PRL84},
resistivity along such direction, $R_{[100]}$ ($R_{[110]}$), should be equal to another resistivity, $R_{[010]}$ ($R_{[1\bar{1}0]}$). 
Above the EN ordering field $B^\star \sim 30 T$, however, the anisotropy between $R_{[100]}$ ($R_{[110]}$) and $R_{[010]}$ ($R_{[1\bar{1}0]}$) has been observed indicating the $B_\mathrm{1g}$ ($B_\mathrm{2g}$) symmetry breaking of electronic states.
Crystal symmetry breaking (CSB) accompanied by the EN ordering with the frequency change of de Haas-van Alphen (dHvA) oscillations 
\cite{Jiao_PNAS112}
has also been indicated by magnetostriction and ultrasonic measurements
\cite{Rosa_PRL122, Kurihara_PRB101}.

In CeRhIn$_5$, CSB due to the electric multipole degree of freedom cannot be expected because the ground state and first and second excited states are described by the three Kramers doublets, $\Gamma_7$, $\Gamma_7$, and $\Gamma_6$
\cite{Willers_PRB81}.
Therefore, there is new physics to be the origin of the FI-EN state.
One of the proposed mechanisms is the enhancement of the two-dimensionality of electronic states due to magnetic fields.
In CeRhIn$_5$, the two-dimensionality of the ground state wave functions in zero magnetic fields is stronger than other related compounds of CeIrIn$_5$ and CeCoIn$_5$
\cite{Willers_PNAS112, Haule_PRB81}.
In addition to this property, magnetic fields enhance the in-plane $p$-$f$ hybridization between Ce-$4f$ and In-$5p$ electrons because the weight of $| J_z = \pm 5/2 >$ with two-dimensional donut shape in the ground state wave functions increases due to Zeeman effect
\cite{Rosa_PRL122}.

To confirm the mechanism of the EN ordering, the high-magnetic-field investigation in the related compound of CeRhIn$_5$ is important in terms of the dimensionality of electronic states.
In CeIrIn$_5$, the quantum states are also described by three Kramers doublets, $\Gamma_7$, $\Gamma_7$, and $\Gamma_6$
\cite{Willers_PRB81}.
In contrast to CeRhIn$_5$, Ce-$4f$ electrons show itinerant behavior due to the Kondo effect at low temperatures
\cite{Shishido_JPSJ71}
and the three-dimensional electronic state has been observed in CeIrIn$_5$
\cite{Willers_PNAS112, Haule_PRB81}.
In CeIrIn$_{5}$, the metamagnetic behavior and a distinct anomaly have been observed in the specific heat, magnetization curve, and magnetic torque at low temperatures and high-magnetic fields for $B//[001]$
\cite{Kim_PRB65, Takeuchi_JPSJ70, Matsuda_JPSJ85, Hirose_JPSCP29, Capan_PRB80}.
The FI Lifshitz transition has also been proposed as the origin of this anomaly because of the frequency change of dHvA oscillations
\cite{Aoki_PRL116}.
Therefore, it is important to investigate the electronic structure in high-magnetic fields in CeIrIn$_5$ in terms of the FI Lifshitz and metamagnetic transitions induced by the EN ordering.
These investigations in CeIrIn$_5$ can provide new findings to understand the EN state in CeRhIn$_5$.

\begin{table*}[htbp]
\caption{
Symmetry strains, electric quadrupoles, and elastic constants corresponding to the irreducible representation (irrep) in $D_{4h}$.
}
\label{table1}
\begin{tabular}{cccc}
\hline
\textrm{IR} 
& \textrm{Symmetry strain}
	& \textrm{Electric quadrupole}
		& \textrm{Elastic constant}
\\ 
\hline
$A_\mathrm{1g}$
	& $\varepsilon_\mathrm{B} = \varepsilon_{xx} + \varepsilon_{yy} + \varepsilon_{zz}$
		& 
			& $ C_\mathrm{B} = \left( 2C_{11} + 2C_{12} + 4C_{13} + C_{33} \right)/9 $
\\

	& $\varepsilon_u = (2\varepsilon_{zz} - \varepsilon_{xx} - \varepsilon_{yy})/\sqrt{3}$
		& $O_{3z^2-r^2} = \left( 3z^2-r^2 \right)/r^2$
			& $C_u = \left( C_{11} + C_{12} - 4C_{13} + 2C_{33} \right)/6$
\\
$B_\mathrm{1g}$
	& $\varepsilon_{x^2-y^2} = \varepsilon_{xx} - \varepsilon_{yy}$
		& $O_{x^2-y^2} = \left( x^2 -y^2 \right)/r^2$
			& $C_\mathrm{T} = \left( C_{11} - C_{12} \right)/2$
\\
$B_\mathrm{2g}$
	& $\varepsilon_{xy}$
		& $O_{xy} =  xy/r^2$
			& $C_{66}$
\\
$E_\mathrm{g}$
	& $\varepsilon_{yz}$
		& $O_{yz} = yz/r^2$
			& $C_{44}$
\\

	& $\varepsilon_{zx}$
		& $O_{zx} = zx/r^2$
			&$C_{44}$ 
\\
\hline
\end{tabular}
\end{table*}

To elucidate the symmetry breaking due to EN ordering, we performed ultrasonic and magnetocaloric effect (MCE) measurements.
The ultrasonic measurement is a powerful tool to observe a phase transition related to the CSB because ultrasonic waves can induce the strains that belong to the irreducible representation (irrep) as listed in Tab. \ref{table1}
\cite{Luthi_PhysAc}.
In addition, we can reveal the electric multipole as an order parameter of such CSB in terms of the multipole-strain interaction.
The MCE is also important to investigate metamagnetic transition in $4f$-electron systems
\cite{Aoki_JMMM177}.

In this study, we investigated the magnetic-field dependence of the elastic constants of CeIrIn$_5$ using ultrasonic measurements and a pulsed magnet.
We revealed that the anisotropic electronic ordering with CSB could be absent in both the metamagnetic and FI Lifshitz transitions.
On the other hand, the charge degree of freedom that is coupled with the isotropic in-plane strain can be the origin of metamagnetic transition.

\section{Experimental procedures}

Single crystals of CeIrIn$_5$ were grown by the flux method.
Sample surfaces perpendicular to the crystallographic orientations were determined using the Laue backscattering method.
The ultrasonic pulse-echo method with a numerical vector-type phase detection technique was used for the ultrasonic velocity $v$
\cite{Fujita_JPSJ80}.
Piezoelectric transducers using LiNbO$_3$ plates with a 36$^\circ$ Y-cut and an X-cut (YAMAJU CO) were employed to generate longitudinal ultrasonic waves with the fundamental frequency of approximately $f = 30$ MHz and the transverse waves with 18 MHz, respectively. 
The elastic constant, $C = \rho v^2$, was calculated from the ultrasonic velocity, $v$, and the mass density, $\rho = 9.188$ g/cm$^3$
\cite{Moshopoulou_APA74}.
The ultrasonic propagation direction, $\boldsymbol{q}$, and the polarization direction, $\boldsymbol{\xi}$, for the elastic constant $C_{ij}$ are shown in the figures in the present paper.
A capacitance thermometer using a non-magnetic ferroelectric KTa$_{1-x}$Ni$_x$O$_3$ was employed for magnetocaloric effects (MCE)
\cite{Miyake_RSI91}.
For high-field measurements up to 46 T, a non-destructive pulse magnet with a time duration of 36 ms installed at The Institute for Solid State Physics, The University of Tokyo, was used.
$^4$He cryostat was used to obtain the low temperatures down to 1.4 K.

\section{Results and discusstions}

\subsection{Elastic anomalies in field-induced transitions}

\begin{wrapfigure}{RB}[0pt]{0.5\textwidth}
\centering
\includegraphics[clip, width=0.5\textwidth, bb=0 -10 330 360]{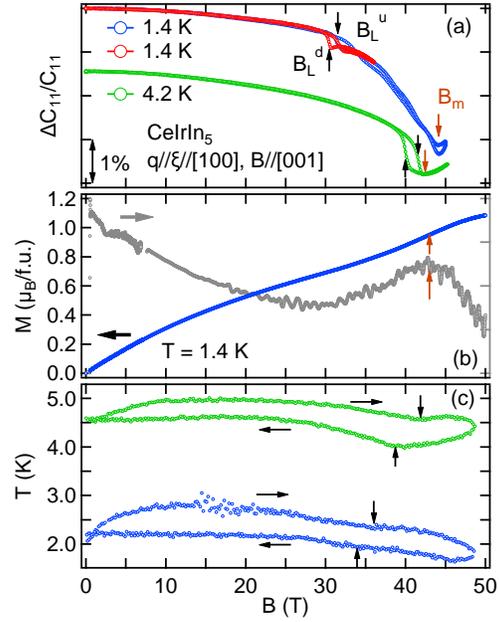}
\caption{
(a) Magnetic-field dependence of the relative variation of the longitudinal elastic constant $\Delta C_{11}/C_{11} = [C_{11}(B) - C_{11}(B = 0)]/C_{11}(B = 0)$ at 1.4 and 4.2 K for $B//[001]$. 
The brown vertical arrows indicate the metamagnetic transition $B_\mathrm{m}$.
The black vertical arrows indicate the FI Lifshitz transition field $B_\mathrm{L}^\mathrm{u}$ ($B_\mathrm{L}^\mathrm{d}$) for field up-sweep (down-sweep).
(b) Magnetization curve and corresponding differential magnetization, $dM/dB$, at 1.4 K for $B//[001]$ in CeIrIn$_5$ measured in the previous study
\cite{Hirose_JPSCP29}.
(c) MCE in CrIrIn$_5$.
The right and left arrows show hysteresis directions.
}
\label{Fig1}
\end{wrapfigure}

To investigate the origin of the FI-EN ordering and the metamagnetic transition, we measured the magnetic-field dependence of elastic constants.
Figure \ref{Fig1}(a) shows the relative variation of the longitudinal elastic constants $\Delta C_{11}/C_{11}$ up to 46 T for $B//[001]$ at 1.4 and 4.2 K.
At 4.2 K, $C_{11}$ exhibits the softening from zero field up to $B_\mathrm{m} = 42$ T with a characteristic hysteresis loop around $B_\mathrm{L}^\mathrm{u} \sim 42$ T for field up-sweep and $B_\mathrm{L}^\mathrm{d} \sim  40$ T for field down-sweep, which are just below  $B_\mathrm{m}$.
At 1.4 K up to 46 T, $C_{11}$ shows the softening up to $B_\mathrm{m} = 44$ T.
We also measured $C_{11}$ up to 36 T to avoid eddy current heating of the sample.
In this measurement, the magnetic field of the elastic anomaly with a hysteresis loop changes from $\sim 40$ T down to $B_\mathrm{L}^\mathrm{u} \sim 32$ T and $B_\mathrm{L}^\mathrm{d} \sim  30$ T. 
Comparing our result to the magnetization curve in the previous study (see Fig. \ref{Fig1}(b))
\cite{Hirose_JPSCP29},
we deduce that $B_\mathrm{m}$ corresponds to the metamagnetic transition.
On the other hand, $B_\mathrm{L}$ can  be consistent with the FI Lifshitz transition
\cite{Aoki_PRL116}.

For further understanding of the FI transitions, we investigate the MCE in CeIrIn$_5$.
Figure \ref{Fig1}(c) shows the magnetic-field dependence of the temperatures for $B//[001]$ measured in the liquid $^4$He condition.
Around $B_\mathrm{L}$, we observed distinct dip structures in the MCE starting from 4.2 K at zero magnetic fields.
We also observed the curvature change in the MCE data from 2.0 K.
The MCE in CeIrIn$_5$ is similar to the dip structure around the metamagnetic transition in CeRu$_2$Si$_2$
\cite{Aoki_JMMM177}.
This is consistent with the previous specific heat measurements, which have revealed the existence of thermodynamic transition under magnetic fields
\cite{Kim_PRB65}.
On the other hand, the MCE results exhibit a hysteresis loop, implying that the eddy current heating causes the temperature change of the sample.
Therefore, the difference between $B_\mathrm{L}^\mathrm{u}$ for field up-sweep and $B_\mathrm{L}^\mathrm{u}$ for down-sweep observed in the ultrasonic measurements originates from the eddy current heating.

\subsection{Investigation of the origin of field-induced Lifshitz and metamagnetic transitions}

\begin{wrapfigure}{RT}[0pt]{0.5\textwidth}
\centering
\includegraphics[clip, width=0.5\textwidth, bb=0 -30 330 360]{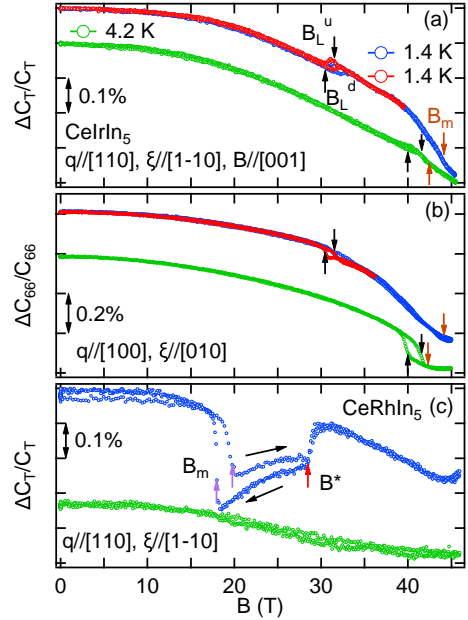}
\caption{
(a) Magnetic-field-dependence of the relative change $\Delta C_{ij}/C_{ij}$ of transverse elastic constants (a) $C_\mathrm{T} =(C_{11} - C_{12})/2$ and (b) $C_{66}$ for $B//[001]$ at 1.4 and 4.2 K. 
(c) Magnetic-field dependence of $\Delta C_\mathrm{T}/C_\mathrm{T}$ in CeRhIn$_5$ for $B//[001]$ at 1.4 and 4.2 K observed in the previous study
\cite{Kurihara_PRB101}.
The purple and red vertical arrows indicate the metamagnetic transition field $B_\mathrm{m}$ and the EN ordering field $B^\star$, respectively.
The right and left arrows show hysteresis directions.
}
\label{Fig2}
\end{wrapfigure}


Our high-magnetic-field measurements showed that both the FI Lifshitz and the metamagnetic transitions distinguished each other.
To discuss the CSB as an origin of these transitions, we measured the transverse elastic constants in CeIrIn$_5$.
We observed the distinct elastic anomaly in the transverse elastic constants. 
Figures $\ref{Fig2}$(a) and $\ref{Fig2}$(b) show the magnetic-field-dependence of the relative change of transverse elastic constants $\Delta C_{ij}/C_{ij}$ for $B//[001]$ at 1.4 and 4.2 K. 
$C_\mathrm{T}$ and $C_{66}$ of CeIrIn$_5$ exhibit the elastic softening from zero magnetic fields up to 46 T with a characteristic hysteresis loop around $B_\mathrm{L}$ and the curvature change around $B_\mathrm{m}$.
Furthermore, each elastic constant shows softening above $B_\mathrm{m}$.

To compare the elastic anomaly in CeRhIn$_5$, we discuss the possibility of the EN ordering with CSB in CeIrIn$_5$.
In CeRhIn$_5$, the transverse elastic constant $C_\mathrm{T} = (C_{11} - C_{12})/2$ with the irrep $B_\mathrm{1g}$ shows the significant anomaly at the EN ordering field $B^\star$ as depicted in Fig. \ref{Fig2}(c)
\cite{Kurihara_PRB101}
while other elastic constants exhibit only small anomalies.
In contrast, we observed the elastic anomalies at $B_\mathrm{L}$ and $B_\mathrm{m}$ in $C_{11}$ with $2A_\mathrm{1g} \oplus B_\mathrm{1g}$, $C_\mathrm{T}$ with the irrep $B_\mathrm{1g}$, and $C_{66}$ with $B_\mathrm{2g}$.
This result indicates that the FI Lifshitz and metamagnetic transitions do not originate from the anisotropic electronic state with CSB, which is described by one active representation.
Furthermore, the largest elastic softening up to $B_\mathrm{m}$ is observed in the longitudinal elastic constant $C_{11}$ consisting of $A_\mathrm{1g}$.
Therefore, we conclude that the origin of the metamagnetic transition is isotropic.

While no anisotropic electronic ordering is observed at $B_\mathrm{L}$ and $B_\mathrm{m}$, the lattice instability can be indicated in high fields in CeIrIn$_5$.
Our high-field ultrasonic measurements shows the softening of $C_\mathrm{T}$ and $C_{66}$ above $B_\mathrm{m}$.
This fact suggests that the crystal instability with the irreps $B_\mathrm{1g}$ and $B_\mathrm{2g}$ remains in high fields.
In other words, the anisotropic electronic state with CBS may be realized in high fields.
Therefore, we deduce that the quadrupole degrees of freedom are activated in high fields due to the enhancement of the in-plane $p$-$f$ hybridization between Ce-$4f$ and In-$5p$ electrons by the Zeeman effect as discussed in CeRhIn$_5$
\cite{Kurihara_PRB101}.
To confirm whether or not the FI-EN ordering appears in CeIrIn$_5$, more high-field ultrasonic measurements are necessary.
This further investigation can provide important information for the contribution of two-dimensionality of the electronic states to the FI-EN ordering in CeRhIn$_5$.

Whereas the FI-EN ordering can be absent in CeIrIn$_5$, intriguing properties in high fields are revealed.
As discussed above, the amount of softening of $C_{11}$ up to $B_\mathrm{m}$ is much larger than others.
This result is similar to the CeRu$_2$Si$_2$ exhibiting significant elastic softening of the longitudinal elastic constants under magnetic fields
\cite{Yanagisawa_JPSJ71}.
Since the strain $\varepsilon_{xx}$ induced by the ultrasonic waves for the longitudinal elastic constant $C_{11}$ consists of the isotropic volume strain $\varepsilon_\mathrm{B}$ with the irrep $A_\mathrm{1g}$ and the tetragonal strain $\varepsilon_u$ with another $A_\mathrm{1g}$, the isotropic charge degree of freedom $A_\mathrm{1g}$ can be a candidate origin of the FI ordering in CeIrIn$_5$.
This high-field elastic property can provide important information to understand the quantum states causing the FI-EN ordering in CeRhIn$_5$ in addition to the metamagnetic and FI Lifshitz transitions in CeIrIn$_5$.


\section{Conclusion}

In conclusion, we investigated the origin of the FI Lifshitz transition and the metamagnetic transition in CeIrIn$_5$ by ultrasonic and MCE measurements.
The longitudinal elastic constant $C_{11}$ exhibited the elastic anomaly at both FI Lifshitz and metamagnetic transitions.
The distinct temperature changes around $B_\mathrm{L}$ observed by the MCE measurements indicated thermodynamic transition.
We also measured the transverse elastic constants to investigate whether or not the FI Lifshitz transition and the metamagnetic transition were described by the anisotropic electronic ordering with the CSB.
However, both $C_\mathrm{T}$ with the irrep $B_\mathrm{1g}$ and $C_{66}$ with $B_\mathrm{2g}$ exhibited the elastic anomalies at $B_\mathrm{L}$ and $B_\mathrm{m}$.
Therefore, we conclude that the EN ordering with the CSB, which has been discussed in CeRhIn$_5$ under magnetic fields, was not the origin of FI Lifshitz transition and metamagnetic transition in CeIrIn$_5$.

\section*{acknowledgement}

The authors thank Yasuhiro H. Matsuda, Hiroshi Yaguchi, and Shusaku Imajo for valuable discussions.
This work was supported by Grans-in-Aid for  young scientists (KAKENHI JP20K14404).
This work was also partly supported by Grans-in-Aid for young scientists (KAKENHI JP22K13999) and scientific research (KAKENHI JP20K03854, JP19K03713, and JP19H00648).

\end{document}